\documentclass[aps,prc,letterpaper,11pt,twoside,tightenlines,nofootinbib,showpacs,preprint,superscriptaddress]{revtex4-1}
\usepackage[colorlinks=true]{hyperref}
\hypersetup{colorlinks=true, citecolor=blue, urlcolor=blue, linkcolor=blue}
\usepackage{graphicx}
\usepackage{epsfig,float}
\usepackage{subcaption}
\usepackage{amsmath}
\usepackage{rotating}   
\usepackage{xcolor}      

\begin{document}
\arraycolsep1.5pt
\newcommand{\Ima}{\textrm{Im}}
\newcommand{\Rea}{\textrm{Re}}
\newcommand{\mev}{\textrm{ MeV}}
\newcommand{\gev}{\textrm{ GeV}}
\newcommand{\red}[1]{\textcolor{red}{#1}}

\title{Searching for bound states in the open strangeness systems}

\author{C. W. Xiao}
\email{xiaochw@gxnu.edu.cn}
\affiliation{Department of Physics, Guangxi Normal University, Guilin 541004, China}
\affiliation{Guangxi Key Laboratory of Nuclear Physics and Technology, Guangxi Normal University, Guilin 541004, China}
\affiliation{School of Physics, Central South University, Changsha 410083, China}

\author{J. J. Wu}
\email{wujiajun@ucas.ac.cn}
\affiliation{School of Physical Sciences, University of Chinese Academy of Sciences (UCAS), Beijing 100049, China}
\affiliation{Southern Center for Nuclear-Science Theory (SCNT), Institute of Modern Physics, Chinese Academy of Sciences, Huizhou 516000, China}

\date{\today}

\begin{abstract}

Inspired by the recent findings of $Z_{cs}$ and $P_{cs}$ states, we investigate the strong interactions of the systems with open strangeness(es) from the light sector to the heavy sector (no beauty quark), where the interaction potential is derived from the vector meson exchange mechanism in $t$- and $u$-channels.
In the current work, we discuss all of single channel cases for the open strangeness in the systemic framework, where the resonances $X_0(2866)$, $D^*_{s0}(2317)$ and $D_{s1}(2460)$ are dynamically generated.
Furthermore, there are many new exotics predicted. 
In addition, the left-hand cut problem in $t$- and $u$-channels is discussed in detail. 

\end{abstract}

\maketitle

\section{Introduction}

Searching for exotic states is a key issue to deeply understand the properties of quantum chromodynamics (QCD), which attract much interest both in theory and experiment. 
In 2021, a new tetraquark-like state $T_{cc}^+$ in the $D^0 D^0 \pi^+$ invariant mass spectrum was discovered by the LHCb Collaboration~\cite{LHCb:2021vvq,LHCb:2021auc}, the mass and the width of which were about 3875.1 MeV and 410 keV, respectively. 
This state looks like a doubly charmed exotic state with spin-parity $J^P = 1^+$ and constituent quarks $cc \bar{u} \bar{d}$, and catches lots of theoretical attention~\cite{Meng:2021jnw,Agaev:2021vur,Ling:2021bir,Chen:2021vhg,Feijoo:2021ppq,Yan:2021wdl,Dai:2021wxi,Weng:2021hje,Xin:2021wcr,Fleming:2021wmk,Ren:2021dsi,Hu:2021gdg,Albaladejo:2021vln,Abreu:2021jwm,Karliner:2021wju,Du:2021zzh,Ortega:2022efc,Wang:2024vjc,Guo:2023xyf}. 
Note that the mass of this state is very close to the first candidate of heavy exotic state $X(3872)$, found by the Belle Collaboration in 2003~\cite{Belle:2003nnu}, and the first charge charmonium-like state $Z_c (3900)$, reported by the BESIII and Belle Collaborations~\cite{BESIII:2013ris,Belle:2013yex} in 2013. 
Especially, the $X(3872)$ is very close to the $D\bar{D}^*$ threshold, less than 0.1 MeV compared to the $D^0 \bar{D}^{*0}$ threshold, and is called $\chi_{c1} (3872)$ now~\cite{pdg2022} for its $J^{PC} = 1^{++}$. 
But, its exotic properties for a non-conventional $q\bar{q}$ state are still under debate, such as its molecular nature~\cite{Tornqvist:2004qy,Swanson:2003tb,Dong:2008gb,Gamermann:2009fv,Gamermann:2009uq,Guo:2014taa,Song:2023pdq,Wang:2023ovj}, see more discussions in the reviews of Particle Data Group~\cite{pdg2022} and Ref.~\cite{Guo:2017jvc}.
For the $Z_c (3900)$, also around the $D \bar{D}^*$ threshold, a challenge issue is to see if it is a molecular state~\cite{Guo:2013sya,Wang:2013daa,Aceti:2014uea,Dong:2021juy}, a tetraquark candidate~\cite{Dias:2013xfa,Braaten:2013boa,Wang:2013vex} or a cusp effect~\cite{Wang:2013cya,Swanson:2014tra,Szczepaniak:2015eza,Liu:2015taa}, see more discussions in Ref.~\cite{Voloshin:2013dpa}.
As done in Refs.~\cite{Dong:2020hxe,Dong:2021juy}, Ref.~\cite{Dong:2021bvy} investigated the interactions of double charm systems, some of which with the strangeness, and found many heavy-heavy hadronic molecules, some of them consistent with the experimental findings. 
Furthermore, recently a new structure with similar mass 3872.5 MeV in the $D \bar{D}$ invariant mass distribution was report by the BESIII Collaboration~\cite{BESIII:2024ths}, which is also nearby the $D \bar{D}^*$ threshold.
As discussed in Refs.~\cite{Dong:2020hxe,Dong:2021juy}, many such new found resonant structures have manifestly a feature that the masses of most of them are close to the thresholds of a pair of hadrons, which are possibly caused by an attractive interaction. 
In view of these discoveries, the picture of hadronic molecule arises. 
Although in principle it is possible to have states near threshold corresponding to non molecular states, this occurs with a very high price of having extremely small scattering lengths and huge effective range for the threshold channel, which in the case of the $X(3872)$ and $T_{cc}(3875)$ have been shown to grossly disagree with experiment~\cite{Song:2023pdq,Dai:2023kwv}.
Typically, for $J^p=1^{++}$ sector, a new state $\chi_{c1}(4010)$ was discovered recently by the LHCb Collaboration~\cite{LHCb:2024vfz}, which indicates that the main component of $X(3872)$ should be $D \bar{D}^*$ while the charmonium state of $\chi_{c1}(2p)$ is possibly the $\chi_{c1}(4010)$~\cite{Wang:2023ovj}.

Now let us turn to the strangeness sector with the light quark replaced by the the strangeness one, a new structure near the $\bar{D}_s D^* / \bar{D}_s^* D$ thresholds, with the mass and the width as about 3982.5 MeV and 12.8 MeV, respectively, was observed by the BESIII Collaboration in 2020~\cite{BESIII:2020qkh}, which was called as $Z_{cs} (3985)$. 
Conversely, in 2021 the LHCb Collaboration reported two new resonances in the $J/\psi K^+$ invariant mass distribution of the $B^+ \to J/\psi \phi K^+$ decay~\cite{LHCb:2021uow}. 
The first one is named as $Z_{cs} (4000)$, with the mass and the width about 4003 MeV and 131 MeV, respectively, which has a mass analogous to the $Z_{cs} (3985)$ but with much larger width, and the second one $Z_{cs} (4220)$ has the mass about 4216 MeV and the width 233 MeV.
More results on the experimental and theoretical status can be found in Refs.~\cite{Chen:2016qju,Esposito:2016noz,Guo:2017jvc,Olsen:2017bmm,Brambilla:2019esw} for reviews and references therein. 
Besides, the $D_{s0}^*(2317)$ was dynamically generated from the $KD$ coupled channel interactions~\cite{Guo:2006fu,Gamermann:2006nm,Faessler:2007gv,Cleven:2010aw,Guo:2011dd}, whereas, the $D_{s1}(2460)$ from the $KD^*$ coupled channel interactions~\cite{Gamermann:2007fi,Guo:2006rp,Faessler:2007us,Cleven:2010aw,Guo:2011dd}.

In the light quark sector, an exotic tetraquark state with $ud \bar{s}\bar{s}$ and $J^P = 1^-$ was predicted around 1.6 GeV in Ref.~\cite{Burns:2004wy}, which could decay to $K^+ K^0$. 
Analogously, an axial-vector isoscalar tetraquark state about 1.4 GeV with $ud \bar{s}\bar{s}$ and $J^P = 1^+$ was claimed in Ref.~\cite{Kanada-Enyo:2005gga} within a flux-tube quark model, which was near the $K K^*$ threshold and could decay to the $K^+ K^+ \pi^-$ final states, and similar predictions were obtained in Ref~\cite{Cui:2005az} with the constituent quark model. 
Using a QCD sum rule, a tetraquark state of $ud \bar{s}\bar{s}$ with $J^P = 0^+$ and isospin $I=1$ was obtained around 1.5 GeV in Ref.~\cite{Chen:2006hy}. 
Furthermore, using the non-relativistic constituent quark model, a possible tetraquark candidate below the $K^* K^*$ threshold was predicted in Ref.~\cite{Wang:2007kb}, with also the $ud \bar{s}\bar{s}$ configuration but with $J^P = 1^+$ and $I=0$, and an analogous result but below the $K K^*$ threshold was found in Ref.~\cite{Gao:2012zza} with the chiral quark model. 
Thus, Ref.~\cite{Liu:2008ck} suggested to search these tetraquark candidates (one can call them as $T_{ss}$ states) of $J^P = 0^+,\ 1^-,\ 1^+$ in the reaction $p  n \to \Lambda \Lambda X(ud \bar{s}\bar{s}) \to \Lambda \Lambda K^+ K^0$ or $\Lambda \Lambda K K^*$.
As we known, there is not much experimental results for the $T_{ss}$-like states, which is the motivation for us to bring more theoretical information to the experiments.

In summary, for the open strangeness system, there may exist many hadronic molecular candidates.
However, a systematic study of these states is still missing. 
In the present work, we focus on the systems with open strangeness both in the light and heavy sectors (but no beauty quark) to look for the molecular candidates near the thresholds of certain channels. 
Our paper is organized as follow. 
In the next section, the formalism of local hidden gauge is presented. 
Then, our results are shown in the following section. 
Finally, a short conclusion is made.

\section{Formalism}

\subsection{Diagrams of the interactions}

\begin{figure}[htbp]
	\begin{subfigure}{0.45\textwidth}
		\centering
		\includegraphics[width=1\linewidth]{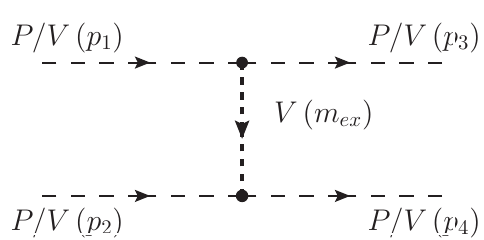} 
		\caption{$t$-channel}
		\label{fig:hidd1}
	\end{subfigure}
	\quad
	\quad
	\begin{subfigure}{0.45\textwidth}  
		\centering 
		\includegraphics[width=1\linewidth]{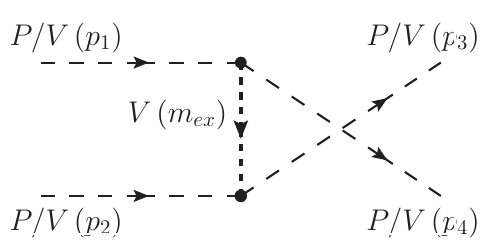}
		\caption{$u$-channel}
		\label{fig:hidd2}  
	\end{subfigure}	
	\caption{Diagrams for the $t$- and $u$-channels with vector meson exchanged mechanisms.}
	\label{fig:hidden}
\end{figure}

\begin{figure}[htbp]
\centering
	\includegraphics[width=0.45\linewidth]{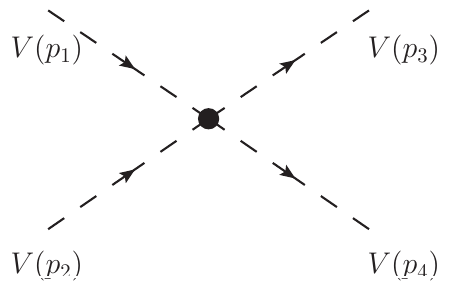} 
	\caption{Diagrams for the contact term.}
	\label{fig:contact}
\end{figure}

In this work, we will consider the following diagrams for the hadron-hadron interaction of $t$- and $u$-channels, as shown in Fig.~\ref{fig:hidd1} and ~\ref{fig:hidd2}, respectively,
where $V$ stands for the vector meson, $P$ represents the pseudoscalar meson, and the particles $V$ and $P$ are specified with their corresponding momenta $p_i$ ($i=1,2,3,4$) or the mass of the exchanged vector meson $m_{ex}$. 
For the $t$- and $u$-channels, the vector meson exchange mechanism is taken into account based on the local hidden gauge formalism~\cite{Bando:1984ej,Bando:1987br,Meissner:1987ge}. 
Note that, we do not consider the $s$-channel by exchanging vector meson, 
since it leads to a $p$-wave amplitude for the $VV$ interactions under the constraint $L + S + I=$ even~\cite{Molina:2008jw} and does not have any exchange meson in some open strangeness and charm systems.
In addition, for the $VV$ interactions, one should also take into account the contribution of the contact term in the tree level, as shown in Fig.~\ref{fig:contact}. 
In fact, from the local hidden gauge Lagrangians, there is no contact term for the $PP$ and $PV$ interactions, which are well described by the chiral Lagrangians, see the one used in Ref.~\cite{Roca:2005nm} for the $PV$ interaction as discussed below. 
Of course, there is no $PPP$ vertex in the local hidden gauge formalism, due to angular momentum and parity conservation. 


\subsection{Interaction Lagrangians}

The local hidden gauge Lagrangians involving the vertexes of the exchanged vector mesons are given by~\cite{Bando:1984ej, Bando:1987br, Meissner:1987ge}
\begin{eqnarray}
{\cal L}_{VVV} &=& ig ~\langle [V_\nu,\partial_{\mu}V_\nu]V^{\mu}\rangle, \label{eq:vvv} \\
{\cal L}_{VPP} &=& -ig ~\langle [P,\partial_{\mu}P]V^{\mu}\rangle, \label{eq:vpp} 
\end{eqnarray}
and for the contact term of $VV$ interaction, the Lagrangian is written as ~\cite{Bando:1984ej,Bando:1987br,Meissner:1987ge}
\begin{equation}
{\cal L}_{VVVV} = \frac{g}{2} \langle~ [V_\mu,~ V_\nu]~ V^{\mu} V^\nu\rangle, \label{eq:vvvv} 
\end{equation}
where the coupling is $g=m_V/(2f)$ with the pion decay constant $f=93$~MeV and taking $m_V = m_\rho$.
$P$ and $V_\mu$ are the SU(4) matrix of the pseudoscalar and vector meson fields in the Lagrangian terms, respectively.
From Eqs.~\eqref{eq:vvv} and~\eqref{eq:vpp}, one can see that there is a minus sign different from them, which will lead to the same structure for the potentials of the $PP$, $VP$, and $VV$ interactions by the extra factor $\epsilon \cdot \epsilon^\prime = - \vec{\epsilon} \cdot \vec{\epsilon}\,^\prime = -1$ when taking the approximations of the external vector with $\epsilon^0=0$ and $\vec{p}_V=0$ (already implicit in Eq.~\eqref{eq:vvv}). 
Thus, we can make a general formalism for the interactions of $PP$, $VP$, and $VV$ with the potentials derived from Eq.~\eqref{eq:vvv} and ~\eqref{eq:vpp}, see more discussions later. 
Note that, the formalism used in Ref.~\cite{Roca:2005nm} is a bit different from the normal coupled channel Bethe-Salpeter equation, see the discussions later.

\subsection{Derived potentials}

Applying the vector meson exchange mechanism for the systems with open strangeness(es), the interaction potentials are given in Table~\ref{tab:res1}, where we define the potential notations with the exchange meson as
\begin{equation}
V_{t(u)}^{ex} = \frac{-V_{t(u)}}{t(u) - m_{ex}^2 + i \epsilon} , 
\label{eq:vtu}
\end{equation}
with $V_t = (p_1 + p_3)\cdot (p_2 + p_4) =s-u$ and $V_u = (p_1 + p_4)\cdot (p_2 + p_3)=s-t$ for the $t$- and $u$-channels, respectively. 
One should keep in mind that $p_i \ (i=1,\ 2,\ 3,\ 4)$ are the corresponding four momenta of initial and final states, as shown in Fig.~\ref{fig:hidden}, and $m_{ex}$ is the mass of the exchange vector meson. 
Note that, in Table~\ref{tab:res1}, there is a factor $\vec{\epsilon} \cdot \vec{\epsilon}\,^\prime$ for the $VP \to VP$ transitions, and a factor $\epsilon_1 \cdot \epsilon_3\; \epsilon_2 \cdot \epsilon_4$ and $\epsilon_1 \cdot \epsilon_4\; \epsilon_2 \cdot \epsilon_3$ for the $t$- and $u$-channels, respectively, in the case of the $VV \to VV$ transitions, which have been ignored. 
Furthermore, in the results of Table~\ref{tab:res1} the isospin projection has already been made, and thus, the potentials are specified with different isospins. 

\begin{table}[htb]
     \renewcommand{\arraystretch}{1.3}
     \setlength{\tabcolsep}{0.3cm}
\centering
\caption{Interaction potential of the strangeness(es) systems.}  \label{tab:res1}
\begin{tabular}{|c|c|c|}
\hline\hline
Systems & Coefficients $I=0$ & Coes. $I=1\, (\frac{1}{2})$  \\ 
\hline\hline  
$KK$ & $\frac{1}{2} (- 3V_t^\rho + V_t^\omega + 2V_t^\phi ) g^2$    & $\frac{1}{2} (V_t^\rho + V_t^\omega + 2V_t^\phi ) g^2$     \\
          & $- \frac{1}{2} (- 3V_u^\rho + V_u^\omega + 2V_u^\phi ) g^2$  & $+ \frac{1}{2} (V_u^\rho + V_u^\omega + 2V_u^\phi ) g^2$  \\
\hline
$K K^*$ &  $\frac{1}{2} (- 3V_t^\rho + V_t^\omega + 2V_t^\phi ) g^2$    &  $\frac{1}{2} (V_t^\rho + V_t^\omega + 2V_t^\phi ) g^2$    \\
\hline
$K^*K^*$ &  $\frac{1}{2} (- 3V_t^\rho + V_t^\omega + 2V_t^\phi ) g^2$    &  $\frac{1}{2} (V_t^\rho + V_t^\omega + 2V_t^\phi ) g^2$    \\
                 &  $- \frac{1}{2} (- 3V_u^\rho + V_u^\omega + 2V_u^\phi ) g^2$    &  $+ \frac{1}{2} (V_u^\rho + V_u^\omega + 2V_u^\phi ) g^2$    \\
\hline\hline  
$\bar{K}D$ & $\frac{1}{2} (- 3V_t^\rho + V_t^\omega - 2V_u^{D^*_s}) g^2$    &  $\frac{1}{2} (V_t^\rho + V_t^\omega + 2V_u^{D^*_s}) g^2$   \\
\hline
$\bar{K}D^*$  &   $\frac{1}{2} (- 3V_t^\rho + V_t^\omega) g^2$    &  $\frac{1}{2} (V_t^\rho + V_t^\omega) g^2$   \\
\hline
$\bar{K}^*D$  &    $\frac{1}{2} (- 3V_t^\rho + V_t^\omega) g^2$    &  $\frac{1}{2} (V_t^\rho + V_t^\omega) g^2$   \\
\hline
$\bar{K}^*D^*$ &    $\frac{1}{2} (- 3V_t^\rho + V_t^\omega - 2V_u^{D^*_s}) g^2$    &  $\frac{1}{2} (V_t^\rho + V_t^\omega + 2V_u^{D^*_s}) g^2$   \\
\hline\hline  
$KD$ &     $\frac{1}{2} (- 3V_t^\rho - V_t^\omega) g^2$    &  $\frac{1}{2} (V_t^\rho - V_t^\omega) g^2$      \\
\hline
$KD^*$ &   $\frac{1}{2} (- 3V_t^\rho - V_t^\omega) g^2$    &  $\frac{1}{2} (V_t^\rho - V_t^\omega) g^2$       \\
\hline
$K^*D$ &   $\frac{1}{2} (- 3V_t^\rho - V_t^\omega) g^2$    &  $\frac{1}{2} (V_t^\rho - V_t^\omega) g^2$   \\
\hline
$K^*D^*$ &  $\frac{1}{2} (- 3V_t^\rho - V_t^\omega) g^2$    &  $\frac{1}{2} (V_t^\rho - V_t^\omega) g^2$    \\
\hline\hline  
$KD_s$ &   --   &  $(V_t^\phi  + V_u^{D^*}) g^2$      \\
\hline
$KD_s^*$ &  --   &   $V_t^\phi  g^2$      \\
\hline
$K^*D_s$ & --   &  $V_t^\phi  g^2$      \\
\hline
$K^*D_s^*$ &  --   &  $(V_t^\phi  + V_u^{D^*}) g^2$       \\ 
\hline\hline  
$\bar{D}D_s$ &  --   &  $-V_t^{J/\psi}  g^2$          \\
\hline
$\bar{D}D_s^*$ & --   & $-V_t^{J/\psi}  g^2$         \\
\hline
$\bar{D}^*D_s$ &  --  & $-V_t^{J/\psi}  g^2$    \\
\hline
$\bar{D}^*D_s^*$ &   --   &  $-V_t^{J/\psi}  g^2$      \\
\hline\hline  
$DD_s$ &    --  &   $(V_t^{J/\psi} + V_u^{K^*})  g^2$       \\
\hline
$DD_s^*$ &  --     &  $V_t^{J/\psi}  g^2$       \\
\hline
$D^*D_s$ &   --    &  $V_t^{J/\psi}  g^2$      \\
\hline
$D^*D_s^*$ &   --   &  $(V_t^{J/\psi} + V_u^{K^*})  g^2$         \\
\hline\hline  
$D_s D_s$ &   $(V_t^{J/\psi} + V_t^\phi)  g^2$    &   --         \\
                  &   $+ (V_u^{J/\psi} + V_u^\phi)  g^2$    &   --       \\
\hline
$D_sD^*_s$ & $(V_t^{J/\psi} + V_t^\phi)  g^2$    &  --    \\
\hline
$D^*_s D^*_s$ & $(V_t^{J/\psi} + V_t^\phi)  g^2$   &   --      \\
                       &   $+ (V_u^{J/\psi} + V_u^\phi)  g^2$    &   --     \\
\hline\hline
\end{tabular}
\end{table}

In addition, for the $VP \to VP$ transitions, we do not consider the $u$-channel contribution, since this comes from the ``Z" diagram with the Lagrangian ${\cal L}_{VVP}$, the contribution of which was found to be small and could be ignored as discussed in Refs.~\cite{Nakamura:2015qga,Dias:2021upl}. 
\footnote{We keep this problem for further investigation in future when the coupled channel interaction is consider.} 
Besides, for the $VP \to VP$ transitions, a chiral Lagrangian was used in Ref.~\cite{Roca:2005nm}, which is different from the vector meson exchange mechanism taken into account in the present work.
But, the equivalence these two ways was proved by a general derivation of the interaction potential in the appendix of Ref.~\cite{Dias:2021upl}.
Departing from the extrapolated chiral dynamics for the $D\bar{D}$ interaction as done in Ref.~\cite{Gamermann:2006nm}, Ref.~\cite{Bayar:2022dqa} preferred to use the vector meson exchange mechanism for the heavy flavor $PP$ interaction, which is out of the constrain of chiral symmetry, normally suited to the light quark sector.

For the potentials defined in Eq.~\eqref{eq:vtu}, $V_{t(u)}^{ex}$, we need to do the $s$-wave projection, {\it i.e.}, the $t$-channel case, having
\begin{equation}
\begin{aligned}
&\frac{1}{2} \int_{-1}^{+1} \text{d} \cos \theta  \frac{-V_t}{t - m_{ex}^2 + i \epsilon} \equiv \frac{1}{2} \int_{-1}^{+1} \text{d} \cos \theta  \frac{-V_t (s, \cos \theta)}{t(s, \cos \theta) - m_{ex}^2 + i \epsilon}  \\
= & -\frac{s}{\sqrt{\lambda\left(s,m_1^2,m_2^2\right) \lambda\left(s,m_3^2,m_4^2\right)}} 
\left( m_1^2 + m_2^2 + m_3^2+ m_4^2- 2 s - m_{ex}^2  \right)  \\
& \times \log \frac{ m_1^2+m_3^2 - \frac{\left( s+m_1^2-m_2^2\right) \left( s+m_3^2-m_4^2\right)}{2 s} - \frac{\sqrt{\lambda\left(s,m_1^2,m_2^2\right) \lambda\left(s,m_3^2,m_4^2\right)}}{2s} -m_{ex}^2 + i \epsilon}
 {m_1^2+m_3^2 - \frac{\left(s+m_1^2-m_2^2\right) \left(s+m_3^2-m_4^2\right)}{2 s} + \frac{\sqrt{\lambda\left(s,m_1^2,m_2^2\right) \lambda\left(s,m_3^2,m_4^2\right)}}{2s} -m_{ex}^2 + i \epsilon} - 1, \\
 = &- 1 -\frac{s}{\sqrt{\lambda\left(s,m_1^2,m_2^2\right) \lambda\left(s,m_3^2,m_4^2\right)}} 
\left( s_0 - 2 s - m_{ex}^2  \right)  \\
& \times \log \frac{ s \left( s_0 - s - 2m_{ex}^2 \right) - \left(m_1^2-m_2^2\right) \left(m_3^2-m_4^2\right) - \sqrt{\lambda\left(s,m_1^2,m_2^2\right) \lambda\left(s,m_3^2,m_4^2\right)} + i \epsilon}
 {s \left( s_0 - s - 2m_{ex}^2 \right) - \left(m_1^2-m_2^2\right) \left(m_3^2-m_4^2\right) + \sqrt{\lambda\left(s,m_1^2,m_2^2\right) \lambda\left(s,m_3^2,m_4^2\right)} + i \epsilon} ,
\end{aligned}
\label{eq:v2}
\end{equation}
with the K\"all\'en function $\lambda(a,b,c) = a^2 + b^2 + c^2 - 2 (ab + ac + bc)$, $s=(p_1+p_2)^2$, and $s_0 \equiv \left( m_1^2 + m_2^2 + m_3^2+ m_4^2  \right)$, where $m_i \ (i=1,\ 2,\ 3,\ 4)$ are the corresponding masses of initial and final states. For the $u$-channel case, one just takes the replacement $m_3  \leftrightarrow m_4$. 

For the contact term of $VV$ interaction, the interaction $A(1) + B(2) \to C(3) + D(4)$, the amplitude derived from Eq.~\eqref{eq:vvvv} gives rise to three products of polarization vectors in the order of 1, 2, 3, 4, written as
\begin{equation}
\epsilon_\mu \epsilon^\mu \epsilon_\nu \epsilon^\nu, \qquad \epsilon_\mu \epsilon_\nu \epsilon^\mu \epsilon^\nu, \qquad \epsilon_\mu \epsilon_\nu \epsilon^\nu \epsilon^\mu,
\end{equation}
which will lead to the spin components of $J= 0,\ 1,\ 2$ for the amplitude. 
As done in Ref.~\cite{Molina:2008jw}, taking the approximation $\frac{q^2}{M_V^2} = 0$, one can apply the spin projectors to separate the spin components of the amplitude, given by
\begin{equation}
\begin{aligned}
& {\cal P}^{(0)} = \frac{1}{3} \epsilon_\mu \epsilon^\mu \epsilon_\nu \epsilon^\nu, \\
& {\cal P}^{(1)} = \frac{1}{2} \left(  \epsilon_\mu \epsilon_\nu \epsilon^\mu \epsilon^\nu - \epsilon_\mu \epsilon_\nu \epsilon^\nu \epsilon^\mu  \right), \\
& {\cal P}^{(2)} = \left[  \frac{1}{2} \left(  \epsilon_\mu \epsilon_\nu \epsilon^\mu \epsilon^\nu + \epsilon_\mu \epsilon_\nu \epsilon^\nu \epsilon^\mu  \right) - \frac{1}{3} \epsilon_\mu \epsilon^\mu \epsilon_\nu \epsilon^\nu \right],
\end{aligned}
\end{equation}
with the order of 1, 2, 3, 4 for the polarization vectors, which, in fact, is analogous to the isospin projection operators of two-pion system~\cite{Holz:2022smu}, and different from the method proposed in Ref.~\cite{Gulmez:2016scm}. 
Thus, with these spin projectors, we obtain the results of the spin projection to the $VV$ interaction amplitudes of the contact term as shown in Table~\ref{tab:res2}, which are consistent with the results of Ref.~\cite{Molina:2010tx}.

Furthermore, taking these spin projectors, one can easily find that the spin components $J= 0,\ 1,\ 2$ of the $VV$ interaction amplitudes with vector meson exchanged are the same for the $t$-channel with the structure of the polarization vectors $\epsilon_1 \cdot \epsilon_3\; \epsilon_2 \cdot \epsilon_4$, whereas, only the ones with spin $J= 0,\ 2$ are equal for the $u$-channel with the structure of the polarization vectors $\epsilon_1 \cdot \epsilon_4\; \epsilon_2 \cdot \epsilon_3$, since the one with  spin $J= 1$ has a minus sign.

\subsection{Scattering equation}

The scattering amplitude of the coupled channel interaction is evaluated from the coupled channel Bethe-Salpeter equation with the on-shell description, given by~\cite{Oller:1997ti,Oset:1997it}
\begin{equation}
T = [1-VG]^{-1}V ,  \label{eq:BS}
\end{equation}
where the $V$ matrix is made from the transition potentials as discussed above, and the diagonal $G$ matrix is constructed by the loop functions of two intermediate mesons in a certain channel. 
Note that, as discussed above for the $VP$ interactions, a different form of the Bethe-Salpeter equation was taken in Ref.~\cite{Roca:2005nm} due to the polarization vectors appearing in the potential, where a slight correction $1 + \frac{1}{3} \frac{q^2}{M_V^2}$ was also introduced to the loop function of the vector-meson propagator. 
As discussed before, the transition potentials shown in Table~\ref{tab:res1} for the $VP$ interactions in fact have a ignored factor $\vec{\epsilon} \cdot \vec{\epsilon}\,^\prime$, which means that a minus sign from $\epsilon \cdot \epsilon^\prime = - \vec{\epsilon} \cdot \vec{\epsilon}\,^\prime$ is already added to the potentials, see the discussions after Eq.~\eqref{eq:vpp}. 
Besides, as mentioned above, the approximation $\frac{q^2}{M_V^2} = 0$ is taken for the $VV$ interaction potential, and thus, taking the same approximation, the corrected term $\frac{1}{3} \frac{q^2}{M_V^2}$ is indeed small~\cite{Roca:2005nm} and can be ignored safely in the loop function of the $VP$ interaction. 
Therefore, for the consistency of our formalism, the general form of Eq.~\eqref{eq:BS} is applied for all three cases of the interactions, $PP$, $VP$, and $VV$.
The element of $G$ matrix in the $i$-th channel is given by
\begin{equation}
 G _ { i i } ( s ) = i \int \frac { d ^ { 4 } q } { ( 2 \pi ) ^ { 4 } } \frac { 1 } { q ^ { 2 } - m _ { 1 } ^ { 2 } + i \varepsilon } \frac { 1 } { \left( p _ { 1 } + p _ { 2 } - q \right) ^ { 2 } - m _ { 2 } ^ { 2 } + i \varepsilon }  \text{ ,}
 \label{eq:loopf}
\end{equation}
where $p_{1}$ and $p_{2}$ are the four-momenta of the two initial states, respectively, and $m_{1}$, $m_{2}$ the masses of the two intermediate particles for the $i$-th channel appearing in the loop. 
It should be mentioned that the $G$ function is logarithmically divergent. 
To solve this singular integral, one either uses the three-momentum cutoff method~\cite{Oller:1997ti}, where the analytic expression is given by Refs.~\cite{Oller:1998hw,Guo:2005wp}, or takes the dimensional regularization method~\cite{Oller:2000fj}. 
Utilizing the cutoff method, Eq.~\eqref{eq:loopf} can be rewritten as~\cite{Oller:1997ti}
\begin{equation}
G _ { ii } ( s ) = \int _ { 0 } ^ { q _ { \max } } \frac { q ^ { 2 } d q } { ( 2 \pi ) ^ { 2 } } \frac { \omega _ { 1 } + \omega _ { 2 } } { \omega _ { 1 } \omega _ { 2 } \left[ s - \left( \omega _ { 1 } + \omega _ { 2 } \right) ^ { 2 } + i \varepsilon \right] }   \text{  ,}
\label{eq:gco}
\end{equation}
with $q=|\vec{q}\,|$ and $ \omega _ { i } = \sqrt{\vec { q } ^ { \:2 } + m _ { i } ^ { 2 }} $, where the cutoff $q_{max}$ is the free parameter. 

Moreover, the mass and the decay width of the state generated in the coupled channel interaction can be determined just by looking for the pole of the scattering amplitude in the complex Riemann sheets. 
Thus, one needs to extrapolate analytically the scattering amplitude in the complex energy plane, where the $G$ function should be extrapolated to the second Riemann sheet by the continuity condition, given by
\begin{equation}
  G_{i i}^{(II)}\left(s\right) =G_{i i}^{(I)}\left(s\right)-2 i \operatorname{Im} G_{i i}^{(I)}\left(s\right)
  =G_{i i}^{(I)}\left(s\right)+\frac{i}{4 \pi} \frac{p_{cmi} (s)}{\sqrt{s}}  \textbf{ ,}
\end{equation}
for $\text{Re} (\sqrt{s}) > \sqrt{s_{th}},\ \text{Im}\; p_{cmi} > 0$ (see Ref.~\cite{Roca:2005nm}), 
where the loop function of the first Riemann sheet, $G_{i i}^{(I)}(s)$, is given by Eq. \eqref{eq:gco}, and the three momentum in center-of-mass (CM) frame is given by 
\begin{equation}
p_{cmi} (s)=\frac{ \sqrt{\lambda(s, m_1^2, m_2^2)} }{2\sqrt{s}}\, ,
\end{equation}
with the usual K\"all\'en triangle function $\lambda(a,b,c)$, defined above.

\subsection{Discussions for the left-hand cut problem}

The numerator and denominator of the $\log$ in Eq.~\eqref{eq:v2} maybe zero, which is equivalent to $t-m_{ex}^2=0$ at $\cos\theta =\pm 1$.
It is known as left-hand cut.
In fact, we can rewrite $t$ and $u$ as the functions of $s$ and $\cos\theta$, {\it i.e.}, $t\equiv t[s,\cos\theta]$, and thus, the left-hand cut appears at the solution of the constraint $t[s,\cos\theta] \equiv m_{ex}^2$. 
In other word, it is crucial to understand the left-hand cut in view of the analytical behaviours of the function $f[s,\cos\theta] = t[s,\cos\theta] -m_{ex}^2$. 
Then following Ref.~\cite{Lutz:2015lca}, we can define the contour functions $c_{\pm}^t(t^\prime)$ as follows,
\begin{equation}
t[c_{\pm}^t(t^\prime), \pm1]=t^\prime,
\end{equation}
with the well-known roots, given by
\begin{equation}
c_{\pm}^t(t^\prime) =  \frac{1}{2} \left[  s_0 - t^\prime - \frac{1}{t^\prime} \left( m_1^2- m_3^2 \right) \left( m_2^2 - m_4^2 \right) 
\pm \frac{\sqrt{\lambda\left(t^\prime,m_1^2,m_3^2\right) \lambda\left(t^\prime,m_2^2,m_4^2\right)}}{t^\prime}  \right],
\label{eq:croot}
\end{equation}
which is kept to discuss the analytical properties of Eq.~\eqref{eq:v2} later, and similar for the case of $u$-channel. 

Taking the approximation, $t(u) \to 0$, as done in Refs.~\cite{Molina:2008jw,Geng:2008gx}, one can obtain
\begin{equation}
V_{t(u)}^{ex} \to \frac{1}{ m_{ex}^2} V_{t(u)},
\label{eq:v1}
\end{equation}
where the part $V_t$ should be projected to the $s$-wave, having
\begin{equation}
\begin{aligned}
V_t & \to \frac{1}{2} \left[ 3s - \left( m_1^2 + m_2^2 + m_3^2+ m_4^2  \right) - \frac{1}{s} \left( m_1^2- m_2^2 \right) \left( m_3^2 - m_4^2 \right) \right], \\
& =  \frac{1}{2} \left[ 3s - s_0 - \frac{1}{s} \left( m_1^2- m_2^2 \right) \left( m_3^2 - m_4^2 \right) \right]. 
\end{aligned}
\label{eq:vtprj}
\end{equation}
Whereas, for the $u$-channel one just needs to change $m_3 \leftrightarrow m_4$.

\begin{figure}[htb]
\centering
\includegraphics[scale=0.6]{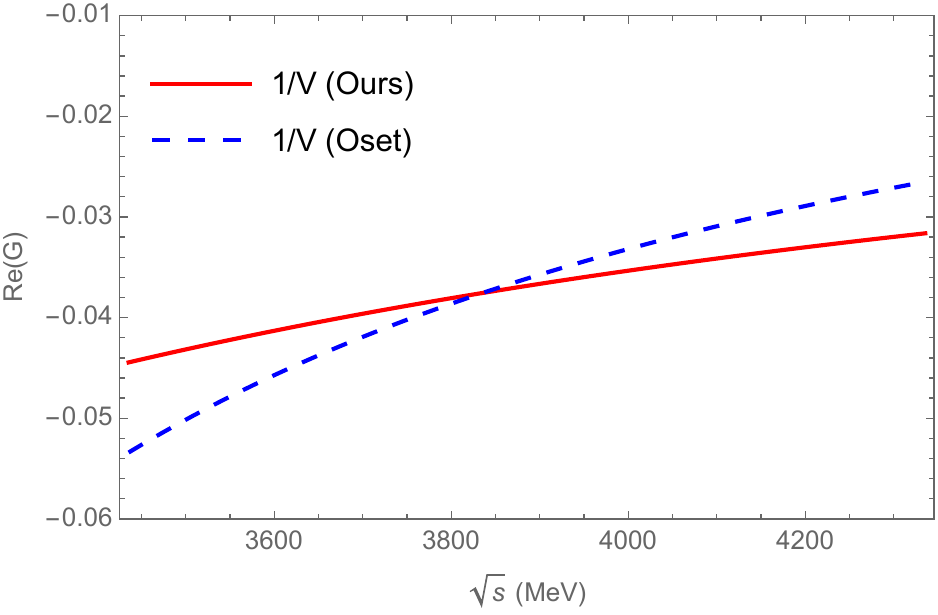}
\caption{Inverse of the Potentials of $\bar{D} D_s \to \bar{D} D_s$ with Eqs.~\eqref{eq:v2} [Ours], \eqref{eq:v1} [Oset].}
\label{fig:Vcom1}
\end{figure}

This approximation is fine for the heavy cases, such as the transition $\bar{D} D_s \to \bar{D} D_s$, see Fig.~\ref{fig:Vcom1} for the comparison of Eqs.~\eqref{eq:v2} and \eqref{eq:v1}.
In fact, in this case, the singularity from the left-hand cut is far away from the threshold.

For the potential of Eq.\eqref{eq:v2}, one should be careful with the $\log$ function.
As found in Ref.~\cite{Wang:2023aza} for the non-diagonal channel of the $V V \to P P$ transitions, a discontinuity appears below the threshold of the bound channel, which in fact is given by the condition 
\begin{equation}
m_1^2+m_3^2 - \frac{\left(s+m_1^2-m_2^2\right) \left(s+m_3^2-m_4^2\right)}{2 s} -m_{ex}^2 \equiv 0, 
\end{equation}
because for this case $\sqrt{\lambda\left(s,m_1^2,m_2^2\right) \lambda\left(s,m_3^2,m_4^2\right)}$ is a purely imaginary and thus the $\log$ function behaves as the $\arctan$ function, see a complicated case in Ref.~\cite{Holz:2015tcg}. 
For the present case of the single channel, which is the diagonal one, we do not face this problem. 
But, for the bound state that we look for, which is located below the threshold and leads to
$\sqrt{\lambda\left(s,m_1^2,m_2^2\right) \lambda\left(s,m_3^2,m_4^2\right)}$ being real, a serious singularity of the left-hand cut will be happened at the energy under the constraint
\begin{equation}
m_1^2+m_3^2 - \frac{\left(s+m_1^2-m_2^2\right) \left(s+m_3^2-m_4^2\right)}{2 s} \pm \frac{\sqrt{\lambda\left(s,m_1^2,m_2^2\right) \lambda\left(s,m_3^2,m_4^2\right)}}{2s} -m_{ex}^2 \equiv 0,
\end{equation}
of which the solutions are given by Eq.~\eqref{eq:croot}, having $s=c_{\pm}^t(m_{ex}^2)$.

\begin{figure}[htb]
\centering
\includegraphics[scale=0.6]{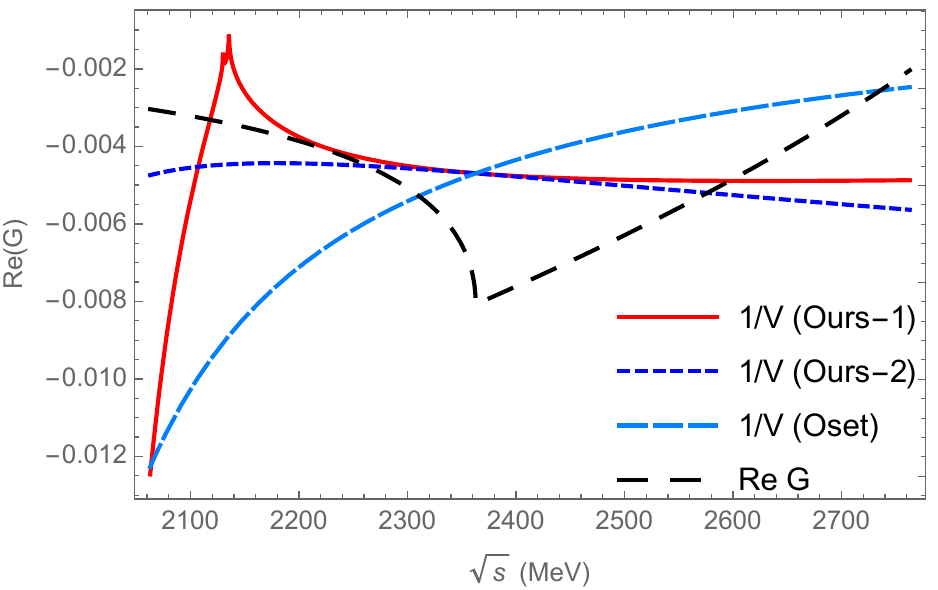}
\caption{Inverse of the Potentials with Eqs.~\eqref{eq:v2} [Ours-1], \eqref{eq:v1} [Oset] and \eqref{eq:v3} [Ours-2] for the channel $K D$ compared with the $G$ function taking $q_{max} = 875$ MeV.}
\label{fig:Vcom}
\end{figure}

If this left-hand cut is just close to the energy region that we are interested in, it will be affected the results seriously as found in Ref.~\cite{Du:2018gyn}. 
In contrast, for the case of the left-hand cut far away from the concerned energy region as discussed in Ref.~\cite{Wang:2022pin}, one can safely ignore them, as discussed above and shown in Fig.~\ref{fig:Vcom1}.
Indeed, the singularity of the left-hand cut in Eq.\eqref{eq:v2} is difficult to normalize with the help of the term $i \epsilon$ due to the fact $\epsilon \to 0$ too. 
And thus, to avoid the singularity, one also can take an improved approximation for Eq.~\eqref{eq:vtu} by ignoring the angle part in the propagator, writing
\begin{equation}
\begin{aligned}
V_{t(u)}^{ex} &\to \frac{-V_{t(u)}}{m_1^2+m_3^2 - \frac{\left( s+m_1^2-m_2^2\right) \left( s+m_3^2-m_4^2\right)}{2 s}  - m_{ex}^2}  \\
& = \frac{-V_{t(u)}}{\frac{1}{2} \left[ \left( s_0 - s \right)  - \frac{1}{s} \left(m_1^2-m_2^2\right) \left( m_3^2-m_4^2\right) \right]  - m_{ex}^2} ,
\end{aligned}
\label{eq:v3}
\end{equation}
with $V_{t(u)}$ projected to the $s$ wave as done in Eq.~\eqref{eq:vtprj}. The compared results of Eqs.~\eqref{eq:v2} [Ours-1], \eqref{eq:v1} [Oset] and \eqref{eq:v3} [Ours-2] are shown in Fig.~\ref{fig:Vcom} for the case of $K D$ channel, which are also compared with the real part of the $G$ function.

Note that, as discussed in Ref.~\cite{Geng:2016pmf}, the singularity of the left-hand cut found in the $\rho \rho$ interaction~\cite{Gulmez:2016scm} was unphysical and the authors avoided the factorization of the potential in the loops by looking at the actual loops with four propagators, where one could define an effective potential from the one-loop level diagrams to obtain the full Bethe-Salpeter series under the ``on-shell'' factorization. 
We have performed the evaluation of the box diagram of the one-loop level diagrams, {\it i.e.} for the $D^* K^*$ channel with two $\rho$, $D^*$ and $K^*$ in the loop,  
and try to extract the ``effective'' potential by extending the way suggested in the appendix of Ref.~\cite{Geng:2016pmf}. 
Due to the contribution (the real part) of the box diagram quite small, as found in Refs.~\cite{Molina:2008jw,Geng:2016pmf}, we do not introduce the ``effective'' potential in our formalism to avoid the confusion, where more details can be found in Ref.~\cite{Geng:2016pmf}. 
We just focus on the following potentials, Eqs.~\eqref{eq:v2}, \eqref{eq:v1} and \eqref{eq:v3}, only contributed from the tree level diagram.

\begin{table}[htb]
     \renewcommand{\arraystretch}{1.3}
     \setlength{\tabcolsep}{0.3cm}
\centering
\caption{The contact terms for the VV interactions.}  \label{tab:res2}
\begin{tabular}{|c|c|c|c|}
\hline
Systems & Qua. Num. & Coefficients $I=0$ & Coes. $I=1,\ (\frac{1}{2})$    \\ 
 &  & Spins $S=0,\ 1,\ 2$ &  Spins $S=0,\ 1,\ 2$    \\ 
\hline
$K^*K^*$  & $S=2,\ C=0$  &  0    &  $-8 g^2, \ 0, \ 4 g^2$ \\
\hline
$\bar{K}^*D^*$ & $S=-1,\ C=1$  & $4 g^2, \ 0, \ -2 g^2$ &  $-4 g^2, \ 0, \ 2 g^2$  \\
\hline
$K^*D^*$ & $S=1,\ C=1$  &  $4 g^2, \ 6 g^2, \ -2 g^2$ & 0  \\
\hline
$K^*D_s^*$ & $S=2,\ C=1$  &  0   &  $-4 g^2, \ 0, \ 2 g^2$    \\
\hline
$\bar{D}^*D_s^*$ & $S=1,\ C=0$  &   0   &  $2 g^2, \ 3 g^2, \ -1 g^2$ \\
\hline
$D^*D_s^*$ & $S=1,\ C=2$  &   0   &   $-4 g^2, \ 0, \ 2 g^2$    \\
\hline
$D^*_s D^*_s$ & $S=2,\ C=2$  &  $-8 g^2, \ 0, \ 4 g^2$    &  0   \\
\hline
\end{tabular}
\end{table}

\section{Results}

Note that, from the potentials listed in Table~\ref{tab:res1}, one can see that some of them are repulsive with positive values, especially for the isospin $I=1$ sectors, which can be easily seen from Eqs.~\eqref{eq:vtu} and \eqref{eq:v1}, and thus, these channels with certain isospin can not lead to any bound states and are out of our concern.
For the $D^* D^*_s$ system with the repulsive potential as shown in Table~\ref{tab:res1}, which is in fact that only for the spin $J=0,\ 2$ when one does the spin projection for the $u$-channel potential as discussed above.
However, since the potential of $J=1$ had a minus sign different from the ones with $J=0,\ 2$ and became attractive, see the results of Ref.~\cite{Molina:2010tx}, a strong cusp around the threshold was found in Ref.~\cite{Dai:2021vgf} for the $D^* D^*_s$ with $J=1$.
The interactions of $K D_s$ and $K^* D_s^*$ are also repulsive as found in Refs.~\cite{Guo:2009ct} and~\cite{Molina:2010tx}, respectively.
Thus, the sectors $(S=2, C=1)$, $(S=1, C=2)$ and $(S=2, C=2)$, and some repulsive channels in the other sectors are not taken into account for searching for bound states.
Furthermore, from Table~\ref{tab:res1}, we can see that the potential of the $KK$ (with $I=0$) channel are slightly attractive due to the $\rho$ exchange term partly cancelled with the contributions of $\omega$ and $\phi$ exchanged, and the ones of the $\bar{D}^{(*)} D_s^{(*)}$ (with $I=\frac{1}{2}$) channels only come from the heavy $J/\psi$ exchange.
Thus, the systems $KK$ (with $I=0$) and $\bar{D}^{(*)} D_s^{(*)}$ (the whole $S = 1$, $C = 0$ sector with $I=\frac{1}{2}$) are too weak to bind a state. 
At last, in the present work, only the $K^{(*)}D^{(*)}$ and $\bar{K}^{(*)}D^{(*)}$ channels with isospin $I=0$ sector may lead to bound states.

As discussed above, there is a free parameter in our formalism, the cutoff $q_{max}$ in the loop function.
Since we do not consider the coupled channels interactions for the systems with open strangeness(es) without beauty quarks, we just take one cutoff for all of the systems to reduce the uncertainties.
We try to determine a proper value of the cutoff by some known resonance(s). 
In the all of systems that we are concerned, there are two systems that generate the well known resonances.
Recalling that, the $D^*_{s0}(2317)$ state was dynamically reproduced in the $KD$ interactions with its coupled channels in Refs.~\cite{Guo:2006fu,Gamermann:2006nm}, in principle it can also be generated from the single $KD$ interaction.
Besides, Ref.~\cite{Molina:2020hde} took the interaction of the $\bar{K}^* D^*$ channel to dynamically generate the $X_0 (2866)$ state, found recently by the LHCb collaboration~\cite{LHCb:2022sfr} and confirmed by the later work~\cite{LHCb:2024vfz}.
Thus, based on these two resonances, $D^*_{s0}(2317)$ and $X_0 (2866)$, we try to determine the value of the cutoff $q_{max}$.

The results are shown in Table~\ref{tab:res3}, where we take Eq.~\eqref{eq:v2} for the evaluation of the potentials and compare the results with Eq.~\eqref{eq:v1}.
From the results of Table~\ref{tab:res3}, we find that it is impossible to use only one cutoff for generating both of the $D^*_{s0}(2317)$ and $X_0 (2866)$ states in this framework.
Thus, two sets of results are shown in Table~\ref{tab:res3} with two different cutoffs to reproduce them one by one, where $q_{max}=875$ MeV is fixed from the mass of the $X_0 (2866)$ state, and $q_{max}=756$ MeV from the mass of the $D^*_{s0}(2317)$ state. 
 Similarly, for the results by using Oset's approximation, see Eq.~\eqref{eq:v1}, two different cutoffs are used, $q_{max}=1100$ MeV and $q_{max}=843$ MeV, for reproducing the $X_0(2866)$ and $D^*_{s0}(2317)$ states, respectively. 
One can see that when we obtain the $D^*_{s0}(2317)$ state from the $K D$ interaction using $q_{max}=756$ MeV, the $D_{s1}(2460)$ is also dynamically generated from the $K D^*$ interaction as found in Refs.~\cite{Gamermann:2007fi, Guo:2006rp, Faessler:2007us, Cleven:2010aw, Guo:2011dd} (see more discussions in recent reviews~\cite{Liu:2024uxn}), where the result with Oset's approximation ($q_{max}=843$ MeV) is consistent with what we have, a pole at 2454.71 MeV compared to 2456.29 MeV.
It should be mentioned that the $D^*_{s2}(2573)$ was identified as a $K^* D^*$ molecule with $J=2$~\cite{Molina:2010tx} for its pole more bound than the ones with $J=0, 1$, which is consistent with the results obtained with Oset's approximation, a pole at 2600.70 MeV or 2732.68 MeV, see Table~\ref{tab:res3}.

Besides, it looks like the results for the $\bar{K}D^*$ interactions do not have stable pole except for the result with Eq.~\eqref{eq:v1} and $q_{max}=1100$ MeV. 
From Table~\ref{tab:res3}, in the case of $q_{max}=875$ MeV, only the systems of $\bar{K}D$ (with $I=0$) and $\bar{K}^*D^*$ (with $I=0$, $J=0,\ 1$) are stably bound, which are consistent with the results of Eq.~\eqref{eq:v1}. 
Our results with $q_{max}=875$ MeV for the $\bar{K}^*D$ (with $I=0$), $\bar{K}^*D^*$ (with $I=0$, $J=2$), and the whole $S = 1,\ C = 1$ sector become abnormal, where the poles of the bound states in the first Riemann sheet for the single channel interaction have a complex width.
The first point of view that we assume is that the poles are already close to the energy region of the unphysical left-hand cut as discussed above and shown in Fig.~\ref{fig:Vcom}, where we try to make further analysis for the reason discussed next. 
The results taking $q_{max}=756$ MeV seem a bit better, where the $\bar{K}^*D$ (with $I=0$) system becomes bound stably in addition to the ones $KD$ and $KD^*$, as the results obtained with Eq.~\eqref{eq:v1}, and the channels $\bar{K}^*D^*$ (with $I=0$, $J=2$), $K^*D$ (with $I=0$) and $K^*D^*$ (with $I=0$ and different $J$) are affected by the unphysical left-hand cut with abnormal poles.  
Indeed, the results with  Eq.~\eqref{eq:v1} are always bound due to the left-hand cut being totally removed.

\begin{table}[htb]
     \renewcommand{\arraystretch}{1.3}
     \setlength{\tabcolsep}{0.1cm}
\centering
\caption{Poles in the first Riemann sheet (1RS) (Unit: MeV). Our results of Eq.~\eqref{eq:v2} are compared with the results of Eq.~\eqref{eq:v1}.}  \label{tab:res3}
\begin{tabular}{|c|c|c|c|c|c|}
\hline\hline
Systems & Ours  & Oset's    &  Ours-2  & Oset's-2    & Exp. \\ 
$I=0\, (\frac{1}{2})$  &  ($q_{max}=875$)  &  ($q_{max}=1100$)   &   ($q_{max}=756$)  &  ($q_{max}=843$)   &  \\ 
\hline\hline  
$\bar{K}D$ (2362.89)  &  2354.97   & 2346.74    &  2362.49   &  2362.82   &    \\
\hline
$\bar{K}D^*$ (2504.20)  &   2284.28 (?)    &  2502.68    &  --   &  --   &      \\
\hline
$\bar{K}^*D$ (2760.86)  &    $2710.32+i 37.23$    &  2689.01    &  2748.04   &  2741.24    &      \\
\hline
$\bar{K}^*D^*$, $J=0$ (2902.17) &   \red{2865.24}     &   \red{2866.38}     &  2898.16   &  2897.02   &  $X_0(2866)$    \\
\hline
$\bar{K}^*D^*$, $J=1$ (2902.17) &   2895.65     &  2859.98     &  2902.10   &  2895.50   &      \\
\hline
$\bar{K}^*D^*$, $J=2$ (2902.17) &   $2821.08-i 103.27$    &  2745.76    &  $2829.93-i 56.36$   &  2837.49   &      \\
\hline\hline  
$KD$ (2362.89)  &  $2223.28 \pm i 22.32$    &   2262.15  &   \red{2317.34}   &   \red{2317.14}   &   $D^*_{s0}(2317)$      \\
\hline
$KD^*$ (2504.20)  &    $2364.55 - i 38.24$   &  2400.22   &  2454.71   &  2456.29    &   $D_{s1}(2460)$     \\
\hline
$K^*D$ (2760.86)  &   $2673.25 - i 183.79$    & 2504.39     &  $2682.17 \pm i 131.15$  &  2620.27  &    \\
\hline
$K^*D^*$, $J=0$ (2902.17)  &  $2851.45-i 153.33$      &  2713.15     &   $2848.10 + i 105.99$    &  2804.40     &    \\
\hline
$K^*D^*$, $J=1$ (2902.17)  &  $2866.56+i 132.57$    &  2748.72     &   $2857.90\pm i 88.87$     &  2826.76     &      \\
\hline
$K^*D^*$, $J=2$ (2902.17)  &  $2795.35-i 199.80$     &  2600.70     &   $2811.04+i 145.12$    &  2732.68     &     $D_{s2}^*(2573)$     \\
\hline\hline
\end{tabular}
\end{table}

Therefore, trying to remove the left-hand cut, we take the approximation form of Eq.~\eqref{eq:v3} for the calculation of the potentials, where the results are given in Table~\ref{tab:res4} compared with the results of Eq.~\eqref{eq:v1} too.
In this approximation, it takes the new tuned cutoffs $q_{max}=883$ MeV and $q_{max}=760$ MeV, for generating the  $X_0 (2866)$ and $D^*_{s0}(2317)$ states, respectively. 
For the case of $q_{max}=883$ MeV, the systems that are obtained abnormal results with Eq.~\eqref{eq:v2} before, except for the ones $KD$ and $KD^*$ (with $I=0$) become stable now as expected, similar results as before are still obtained.
When we check the potentials of Eq.~\eqref{eq:v3} for these systems in detail, we find that the inverse of the potentials with minus slope, see the short-dash (blue) line of Fig.~\ref{fig:Vcom1}, do not cross with the the real part of the loop functions.
Thus, there is no bound pole, which means that the poles we find now with widths in the first Riemann sheet are in fact virtual states. 
For the other case of $q_{max}=760$ MeV, which is just a bit different from the former one $q_{max}=756$ MeV, the results are not much different than before.
Now we know that these abnormal results, except for the ones $KD$ and $KD^*$ (with $I=0$), are in fact the poles of virtual states, which means that these systems could be also bound when we tune the cutoff $q_{max}$ and be searched for the possible bound state in future experiments. 
\begin{table}[htb]
     \renewcommand{\arraystretch}{1.3}
     \setlength{\tabcolsep}{0.1cm}
\centering
\caption{Poles in the first Riemann sheet (1RS) (Unit: MeV). Our results of Eq.~\eqref{eq:v3} are compared with the results of Eq.~\eqref{eq:v1}.}  \label{tab:res4}
\begin{tabular}{|c|c|c|c|c|c|}
\hline\hline
Systems & Ours-new  & Oset's    &  Ours-new2  & Oset's-2    & Exp. \\ 
 &  ($q_{max}=883$)  &  ($q_{max}=1100$)   &   ($q_{max}=760$)  &  ($q_{max}=843$)   &  \\ 
\hline\hline  
$\bar{K}D$ (2362.89)  &  2354.10   & 2346.74    &  2362.40   &  2362.82   &    \\
\hline
$\bar{K}D^*$ (2504.20)  &   --    &  2502.68    &  --   &  --   &      \\
\hline
$\bar{K}^*D$ (2760.86)  &    $2694.17+i 42.94$    &  2689.01    &  2747.40   &  2741.24    &      \\
\hline
$\bar{K}^*D^*$, $J=0$ (2902.17) &   \red{2866.14}     &   \red{2866.38}     &  2897.82  &  2897.02   &  $X_0(2866)$    \\
\hline
$\bar{K}^*D^*$, $J=1$ (2902.17) &   2894.66    &  2859.98     &  2902.06   &  2895.50   &      \\
\hline
$\bar{K}^*D^*$, $J=2$ (2902.17) &   $2813.41-i 135.96$    &  2745.76    &  $2807.55\pm i 66.51$    &  2837.49   &      \\
\hline\hline  
$KD$ (2362.89)  &  2258.22    &   2262.15  &   \red{2317.36}   &   \red{2317.14}   &   $D^*_{s0}(2317)$      \\
\hline
$KD^*$ (2504.20)  &    2391.78   &  2400.22   &  24555.06   &  2456.29    &   $D_{s1}(2460)$    \\
\hline
$K^*D$ (2760.86)  &   $2698.42 - i 237.63$    & 2504.39     &  $2684.60 \pm i 170.50$  &  2620.27  &    \\
\hline
$K^*D^*$, $J=0$ (2902.17)  &  $2874.93+i 192.31$      &  2713.15     &   $2849.62 + i 135.30$    &  2804.40     &    \\
\hline
$K^*D^*$, $J=1$ (2902.17)  &  $2866.91+i 164.04$    &  2748.72     &   $2857.40\pm i 112.55$     &  2826.76     &      \\
\hline
$K^*D^*$, $J=2$ (2902.17)  &  $2824.87+i 259.85$     &  2600.70     &   $2817.27+i 189.09$    &  2732.68     &     $D_{s2}^*(2573)$     \\
\hline\hline
\end{tabular}
\end{table}

\section{Discussions}

Before closing with conclusions, we make some discussions beyond the current formalism. 
In the present framework, we only find that the bound states would appear in the $K^{(*)}D^{(*)}$ and $\bar{K}^{(*)}D^{(*)}$ channel with isospin $I=0$ sector. 
However, there are two limitations, no $\pi$ exchange and no coupled channels. 
Then various interactions will be missing, for example, for the $K^{(*)}K^{(*)}$ channel, $KK^*$ could exchange $\pi$ in the $u$-channel interaction, while the $KD^* \to K^*D$ process could happen through $\pi$ exchange, which will lead to coupled channels effects between the $KD^*$ and $K^*D$ channels.
From the study of the $Z_c(3900)$, we found that the interaction from the vector meson exchange is not enough to bind two hadron systems. 
To generate the $Z_c(3900)$ state from the $D\bar{D}^*$ interaction, except for the vector meson exchange considered, the pseudoscalar and two-pion exchanges were also taken into account in Ref.~\cite{Aceti:2014uea}. 
Here we will make more discussion on the $\bar{D}^{(*)}D_s^{(*)}$ system.

Note that, as discussed in the introduction, in the experiments the states $Z_{cs}(3985)$ and $Z_{cs}(4000)$ are found near the threshold of the $\bar{D}_s^* D_s + \bar{D}_s D_s^*$ channel. 
But, from the interaction of vector meson exchange, where only the heavy $J/\psi$ is allowed, we can not reproduce any bound state in the similar $\bar{D}^{(*)} D_s^{(*)}$ systems.
Indeed, using the same mechanism in Ref.~\cite{Ikeno:2020mra}, they did not find the pole of bound state for the $Z_{cs}(3985)$ and concluded it as a virtual state from a strong cusp structure in the threshold of $\bar{D}_s D^*$, which was a threshold effect. 
Furthermore, using the heavy quark spin symmetry formalism in Ref.~\cite{Yang:2020nrt}, the $Z_{cs}(3985)$ and other predicted $Z_{cs}^*$ states, as the strange molecular partners of the $Z_c(3900)$ and $Z_c(4020)$ states, could be either a virtual state or a resonance. 
Within the one boson exchange model, Ref.~\cite{Yan:2021tcp} reproduced the $Z_{cs}(3985)$ by the interaction potentials from the axial, scalar, isovector, and vector meson exchanges, where analogously the one boson exchange formalism was also applied in Ref.~\cite{Ding:2021igr} to explain the $Z_{cs}(3985)$ state with the light and heavy meson exchanges. 
For this reason, more investigations should be done in the future.
One should keep in mind that in the present work we only consider the single channel interaction with the interaction potential from the vector meson exchange in $t$- and $u$-channels. 

For the further investigation, we will take into account the coupled channels interactions for these systems with attractive potentials and the other interaction dynamics, such as the pseudoscalar meson exchanges. 

\section{Conclusions}

In the present work, under the local hidden gauge formalism, we derive the interaction potential with the vector meson exchange mechanism of the $t$- and $u$-channels, and study the single channel interaction of the systems with open strangeness(es) from the light sector to the heavy one (without beauty quark). 
We face the left-hand cut singularity when we do the $s$-wave projection to the $t$- and $u$-channels potentials, which looks like a unavoided problem for us when the singularities happen to appear at the energy region that we are concerned with. 
Thus, we make a simplified approximation to this left-hand cut problem, where some stable results are obtained.
In our results, we successfully generate the states $X_0(2866)$, $D^*_{s0}(2317)$ and $D_{s1}(2460)$ in the interaction $\bar{K}^* D^*$, $K D$ and $K D^*$, respectively. 
Furthermore, we also find some other bound systems in different sectors, which will be further investigated in our future work.
At last, from the first results with the single channel interaction with the vector meson exchange, some loose bound systems are necessary to include pseudoscalar exchange and coupled channels effects in further study.

\section*{Acknowledgments}

We acknowledge Prof. Eulogio Oset for useful discussions and careful reading of the manuscript, 
and Drs. Mao-Jun Yan and Ming-Zhu Liu for helpful comments. 
One of us, C.W.X., acknowledges the hospitality of the University of Chinese Academy of Sciences where part of this work was done.
This work is supported by 
the Natural Science Foundation (NSF) of Changsha under Grant No. kq2208257, 
the NSF of Hunan province under Grant No. 2023JJ30647, 
the NSF of Guangxi province under Grant No. 2023JJA110076, 
and the National NSF of China under Grant No. 12365019, 12175239, and 12221005,
and also by National Key Research and Development Program of China under Contracts 2020YFA0406400, 
and also by Chinese Academy of Sciences under Grant No. YSBR-101, 
and also by Xiaomi Foundation / Xiaomi Young Talents Program.

\end{document}